\documentclass[aip,pof,numerical,reprint,amsmath,amssymb]{revtex4-1}

\usepackage{graphicx}
\usepackage{dcolumn}
\usepackage{bm}
\usepackage{soul}
\usepackage{color}

\newcommand{\psd}{derived }
\newcommand{\tru}{true }

\newcommand\Froude{\mbox{\textit{Fr}}}    

\begin{document}

\title{Air flow in a collapsing cavity}

\author{Ivo R. Peters}
\affiliation{
Physics of Fluids, Faculty of Science and Technology, MESA+ Research Institute, and J.M. Burgers Centre for Fluid Dynamics, University of Twente\\
P.O. Box 217, 7500 AE Enschede, The Netherlands
}
\affiliation{James Franck Institute, The University of Chicago,\\ Chicago, Illinois 60637, USA.}
\author{Stephan Gekle}
\affiliation{
Physics of Fluids, Faculty of Science and Technology, MESA+ Research Institute, and J.M. Burgers Centre for Fluid Dynamics, University of Twente\\
P.O. Box 217, 7500 AE Enschede, The Netherlands
}
\affiliation{Physikalisches Insitut, Universit\"at Bayreuth, Germany}
\author{Detlef Lohse}
\affiliation{
Physics of Fluids, Faculty of Science and Technology, MESA+ Research Institute, and J.M. Burgers Centre for Fluid Dynamics, University of Twente\\
P.O. Box 217, 7500 AE Enschede, The Netherlands
}
\author{Devaraj van der Meer}
\affiliation{
Physics of Fluids, Faculty of Science and Technology, MESA+ Research Institute, and J.M. Burgers Centre for Fluid Dynamics, University of Twente\\
P.O. Box 217, 7500 AE Enschede, The Netherlands
}

\date{\today}

\begin{abstract}
We experimentally study the airflow in a collapsing cavity created by the impact of a circular disk on a water surface.
We measure the air velocity in the collapsing neck in two ways: Directly, by means of employing particle image velocimetry of smoke injected into the cavity and indirectly, by determining the time rate of change of the volume of the cavity at pinch-off and deducing the air flow in the neck under the assumption that the air is incompressible. We compare our experiments to boundary integral simulations and show that close to the moment of pinch-off, compressibility of the air starts to play a crucial role in the behavior of the cavity. Finally, we measure how the air flow rate at pinch-off depends on the Froude number and explain the observed dependence using a theoretical model of the cavity collapse.
\end{abstract}


\maketitle


\section{Introduction}
The impact of a solid body on a water surface triggers a series of spectacular events: After a splash, if the impact speed is high enough, a surface cavity is formed which pinches off such that a bubble is entrained~\cite{Worthington1900,Bergmann2006,Duez2007}. Right after pinch-off two strong thin jets are formed~\cite{Gekle2009a}, one shooting upwards and one shooting downwards.

An aspect in the impact on liquids that has drawn very little attention is the influence of the accompanying gas phase. When we take into account the inner gas in the detaching air bubble, we find a singularity in the velocity of the inner gas. Assuming any finite flow rate for the gas, the velocity of the gas will diverge because the area that the gas has to flow through goes to zero. Nature has found a way to avoid a true singularity by letting compressibility limit the speed of the air, but nonetheless the air plays an important role in the final shape of the cavity just before pinch-off~\cite{Gordillo2005,Gordillo2006}, and can even reach supersonic speeds~\cite{Gekle2010}.

The main objective of this paper is to understand what determines the gas flow rate in the case of an impacting disc and to obtain insight into the role of compressibility effects in the air. To this end we apply two different approaches: First, we perform volume measurements to determine the flow rate based on continuity, and second we measure the air flow directly by seeding the air with smoke under laser sheet illumination. We compare and extend our experiments with numerical simulations, where we use one- and two-phase boundary integral simulations, sometimes coupled to compressible Euler equations~\cite{Gekle2010a}, to determine the air flow, with and without taking the dynamics of the gas phase into account. {This paper builds partially on a previous publication by Gekle \textit{et al.}\cite{Gekle2010} where it was shown that the gas velocity reaches supersonic speeds and compressibility influences the shape of the cavity in the neck region. Our new results show how the flow rate depends on the disc size and impact speed, for which we derive a scaling law. We experimentally verify the position of the stagnation point that was found numerically before\cite{Gekle2010} and we determine at which moment compressibility starts to play a role in the dynamics of the cavity collapse.}

We have structured this paper as follows: We first give a brief description of the experimental setup in Section~\ref{sec:Setup}.  Section~\ref{sec:Geometric} explains the method of volume measurements, and the results are combined  with numerical simulations. More specifically, we measure how the air flow rate at pinch-off depends on the Froude number and explain the observed dependence using a theoretical model of the cavity collapse. In Section~\ref{sec:Visualization} we perform a direct determination of the air flow velocity by seeding the air with smoke and illuminating with a laser sheet. Subsequently, we compare the results with the velocities that we determined using volume measurements. Finally, in Section~\ref{sec:Compressibility} we discuss in detail when and how compressibility becomes important.

\section{Experimental setup}\label{sec:Setup}
The experimental setup consists of a water tank with a bottom area of 50 cm by 50 cm and 100 cm in height. A linear motor that is located below the tank pulls a disc through the water surface at a constant speed. This disc is connected to the linear motor by a thin rod. The events are recorded with a Photron SA1.1 high speed camera at frame rates up to $20~\text{kHz}$. Our main control parameter is the Froude number, which is defined as the square of  the impact speed $U_0$, nondimensionalized by the disc radius $R_0$ and the gravitational acceleration $g$:
\begin{equation}
    \Froude=\frac{U_0^2}{gR_0}
\end{equation}
\begin{figure}
    \centering
    \includegraphics[width=8cm]{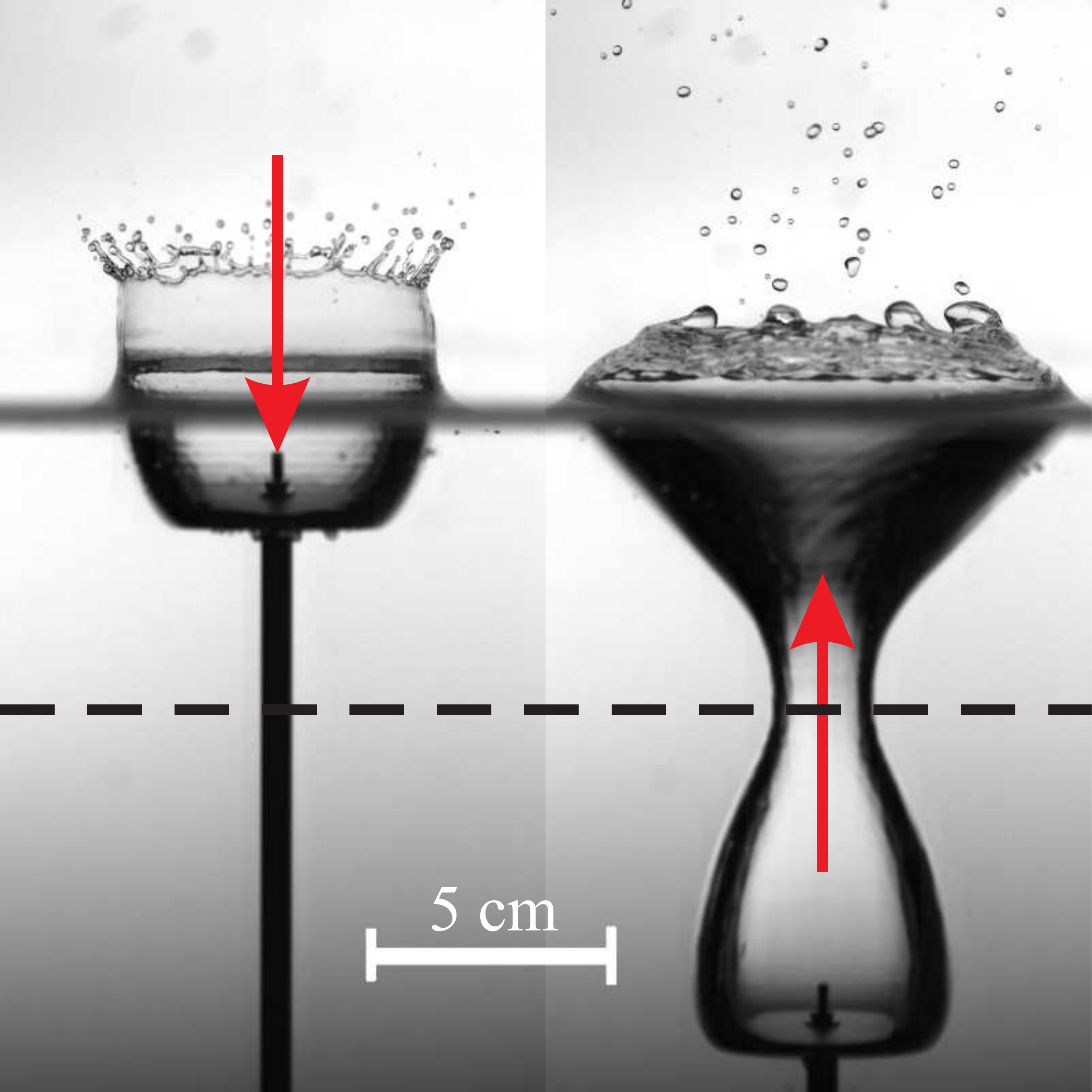}
    \caption{\label{fig:01} Two snapshots of an experiment in which a disk with a radius of $2~\text{cm}$ hits the water surface and moves down at a constant speed of $1~\text{m/s}$. A surface cavity is created that subsequently collapses under the influence of the hydrostatic pressure. Eventually, the cavity pinches off at the depth indicated by the dashed line, and a large air bubble is entrained. The red arrows indicate the direction of the air flow: On the left, volume is created, resulting in a downward air flow. On the right the bubble volume below the pinch-off depth is decreasing, and therefore air is pushed upwards.}
\end{figure}
Two snapshots of the experiment are shown in Fig.~\ref{fig:01}. The left image shows the situation right after impact, when the cavity is formed. A downward flow of air is required to fill in the space that is created by the downward moving disk and the expanding cavity. On the right a later stage in time is shown, some moments before the pinch-off. Here, there is a competition between the downward moving disc and the expanding part of the cavity on the one hand, and the collapsing part, i.e., the region above the maximum, on the other. The former tends to increase the cavity volume below the pinch-off depth (dashed line), whereas the latter decreases it. We always observe that close to pinch-off, the violent collapse is dominant and the bubble volume below the pinch-off point decreases, pushing air out through the neck. As the neck becomes thinner towards the moment of pinch-off, the gas speed increases rapidly. The remaining part of this chapter is devoted to measuring this air flow and comparing the results with numerical simulations.

\section{Geometric approach}\label{sec:Geometric}

The first way in which we will quantify the air flow through the neck of the cavity is an indirect one: We will measure the time evolution of the volume of the cavity below the pinch-off point and calculate its first derivative with respect to time. This will be identified with the air flow rate through the neck. This involves the following assumptions: (i) The air flow is incompressible, (ii) the air flow profile is one-dimensional (i.e., a plug flow) and only directed in the vertical direction, and (iii) the cavity shape is axisymmetric. The first assumption is only violated close to the moment of pinch-off, when the air speed diverges. Compressibility effects at this stage are investigated in \cite{Gekle2010} and its effects will be discussed in section~\ref{sec:Compressibility}. We will justify the second assumption partially by visualizing the air flow inside the cavity and measuring the velocity directly; in addition it is known from two-fluid boundary-integral simulations that the flow profile is very close to one dimensional~\cite{Gekle2010a}. The third assumption only breaks down in the neck-region very close to pinch-off because very small disturbances are remembered during the collapse ~\cite{Doshi2003,Keim2006,Schmidt2009,Turitsyn2009a,Enriquez2010,Enriquez2011,Enriquez2012}. Here, this effect is only relevant locally on a very small scale and can therefore be neglected on the large scale where we measure the volume.

\subsection{Cavity volume}\label{sec:Volume}
We measure the volume of the cavity below the pinch-off depth as illustrated in Fig.~\ref{fig:02}:
\begin{figure}
    \centering
    \includegraphics[width=8cm]{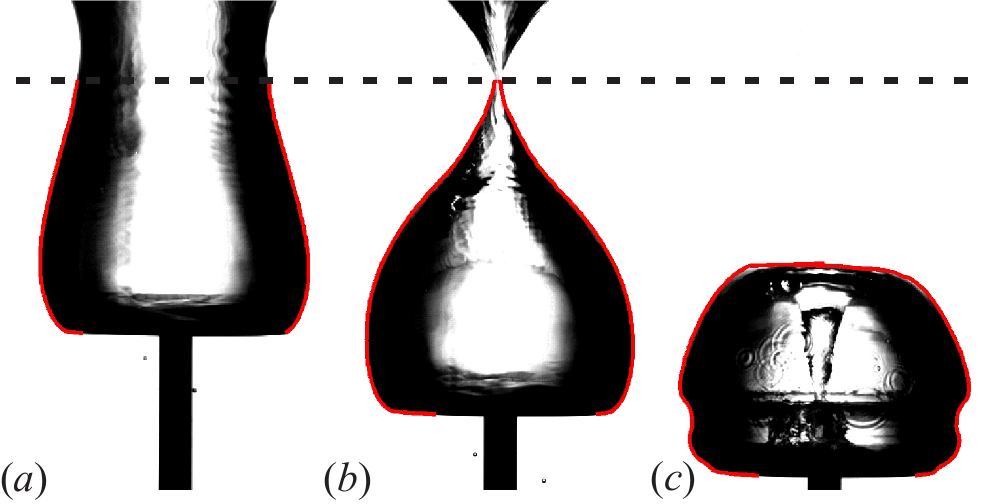}
    \caption{\label{fig:02} (a) The volume of the cavity below the pinch-off depth (dashed line) is determined by tracing the boundary (red line) and assuming symmetry around the central axis. (b) The volume decreases as the neck becomes thinner until the cavity closes. (c) After pinch-off a downward jet enters into the entrapped bubble, and the bubble shows volume-oscillations and cavity ripples.
}
\end{figure}
\begin{figure}
    \centering
    \includegraphics[width=8cm]{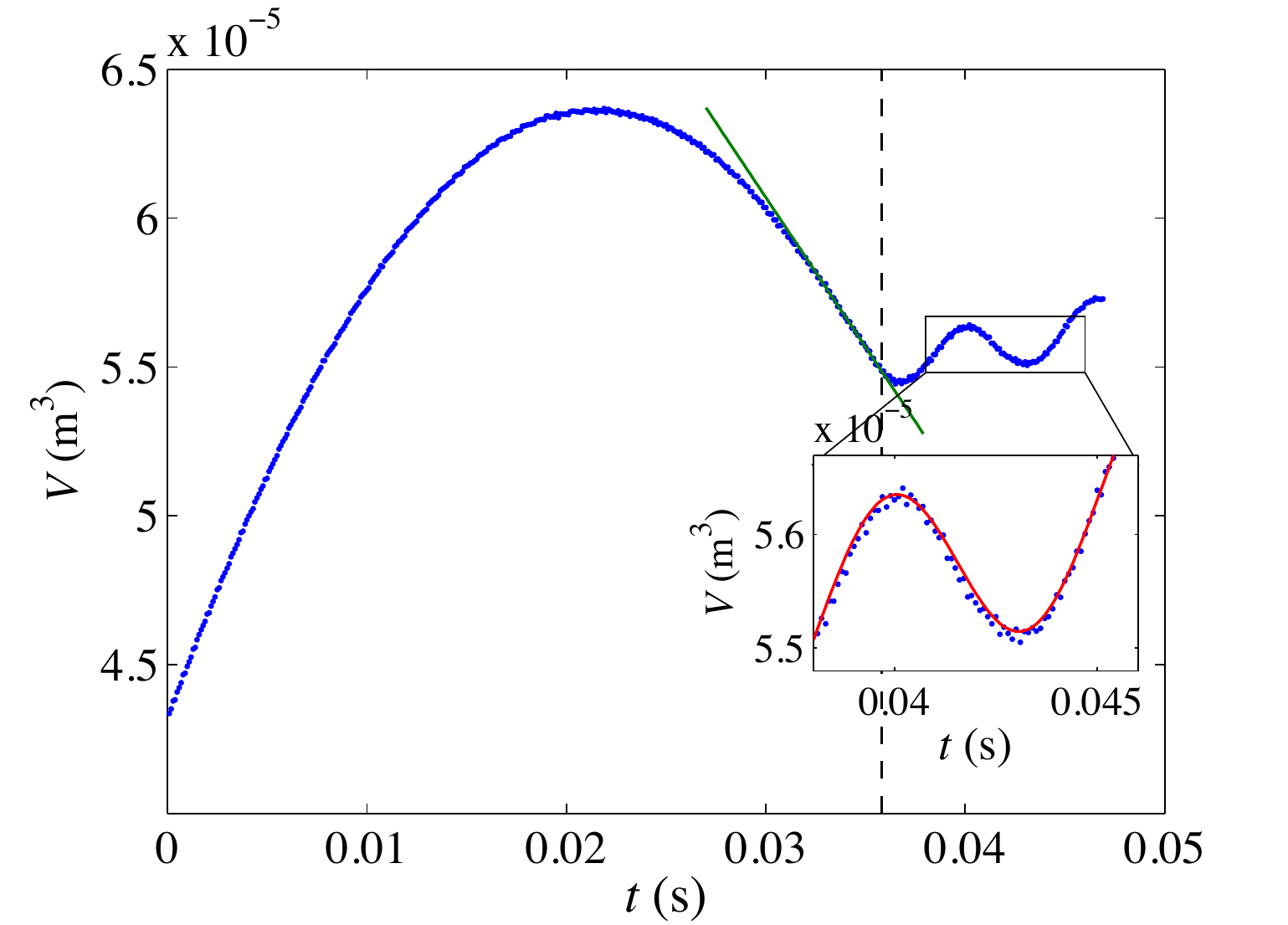}
    \caption{\label{fig:03} Volume below the pinch-off depth as a function of time (blue dots), determined from an experiment with $\Froude=5.1$. The vertical dashed line indicates the moment of pinch-off. Close to pinch-off the volume decrease is well approximated by a linear fit (green line), after pinch-off the bubble oscillates with its resonance frequency (red line: fit with sine function). The steady growth in volume after the pinch-off is caused by the jet entering the bubble, which in our data analysis is not subtracted from the measured bubble volume, see main text.
}
\end{figure}
By tracing the contour for every frame in a movie and invoking axisymmetry we are able to determine the volume of the bubble below the pinch-off depth as a function of time. Such a measurement for a disk with radius $20~\text{mm}$ and impact speed of $1~\text{m/s}$ is shown in Fig.~\ref{fig:03}. In the beginning ($t\lesssim0.022~\text{s}$) the volume is increasing (positive slope), which means that the air at the pinch-off depth is flowing downwards. At the maximum ($t\approx0.022~\text{s}$) the flux through the pinch-off depth is zero, indicating a local stagnation of the flow at this depth. We will study this stagnation point later, in section~\ref{sec:Compressibility}. After this maximum the volume starts to decrease and the flow is directed upwards. This continues until the moment of pinch-off which is indicated by the vertical dashed line in Fig.~\ref{fig:03}. A linear fit (green line) reveals that the flow rate is approximately constant towards the pinch-off moment. More precisely, the linear fit is the time rate of change of the cavity volume at pinch-off, which is equal to minus the maximum value of volume-based flow rate, $\Phi_V \equiv -dV/dt$, under the assumption of incompressibility of the air. We will use this maximum flow rate $\Phi_V$ to compare the flow rates through the neck at different Froude numbers.

After the pinch-off there is a clear oscillation of the volume together with a slow apparent growth of the bubble. The growth is caused by the liquid jet that is entering the bubble (Fig.~\ref{fig:02}c), as the amount of air is fixed after the pinch-off. Since our focus is on the behavior before pinch-off, we chose not to correct the bubble volume {in the image analysis} by subtracting this jet volume. Also, making such a correction would be complicated by the fact that the jet is imaged through the refracting, curved interface of the air bubble. Nevertheless, we determined the frequency of the oscillation by fitting a sine function (red line) after correcting for the slightly positive slope. For the conditions of Fig.~\ref{fig:03} the measured frequency is $143~\text{Hz}${, which is close to the expected resonance --or Minnaert-- frequency\cite{Minnaert1933} of $138~\text{Hz}$\footnote{The Minnaert frequency is given by $f = \frac{1}{2\pi r}\left(\frac{3\gamma p_A}{\rho}\right) = (3.26~\mathrm{m/s})/r$ where $r$ is the bubble radius, $\gamma$ the polytropic exponent, $p_A$ the atmospheric pressure and $\rho$ the density of water. We calculated the equivalent radius of a spherical bubble corresponding to the bubble volume at pinch-off, which equals $5.49\cdot10^{-5}~\text{m}^3$}. This agreement was also noted by~\cite{Grumstrup2007} for the impact of freely falling objects in water.}

\subsection{Air flow rate}
\label{subsec:airflowrate}
A characteristic quantity concerning the gas dynamics in a collapsing cavity is the air flow rate, defined as the volume of air that is being displaced per unit time close to pinch-off.
\begin{figure}
    \centering
    \includegraphics[width=8cm]{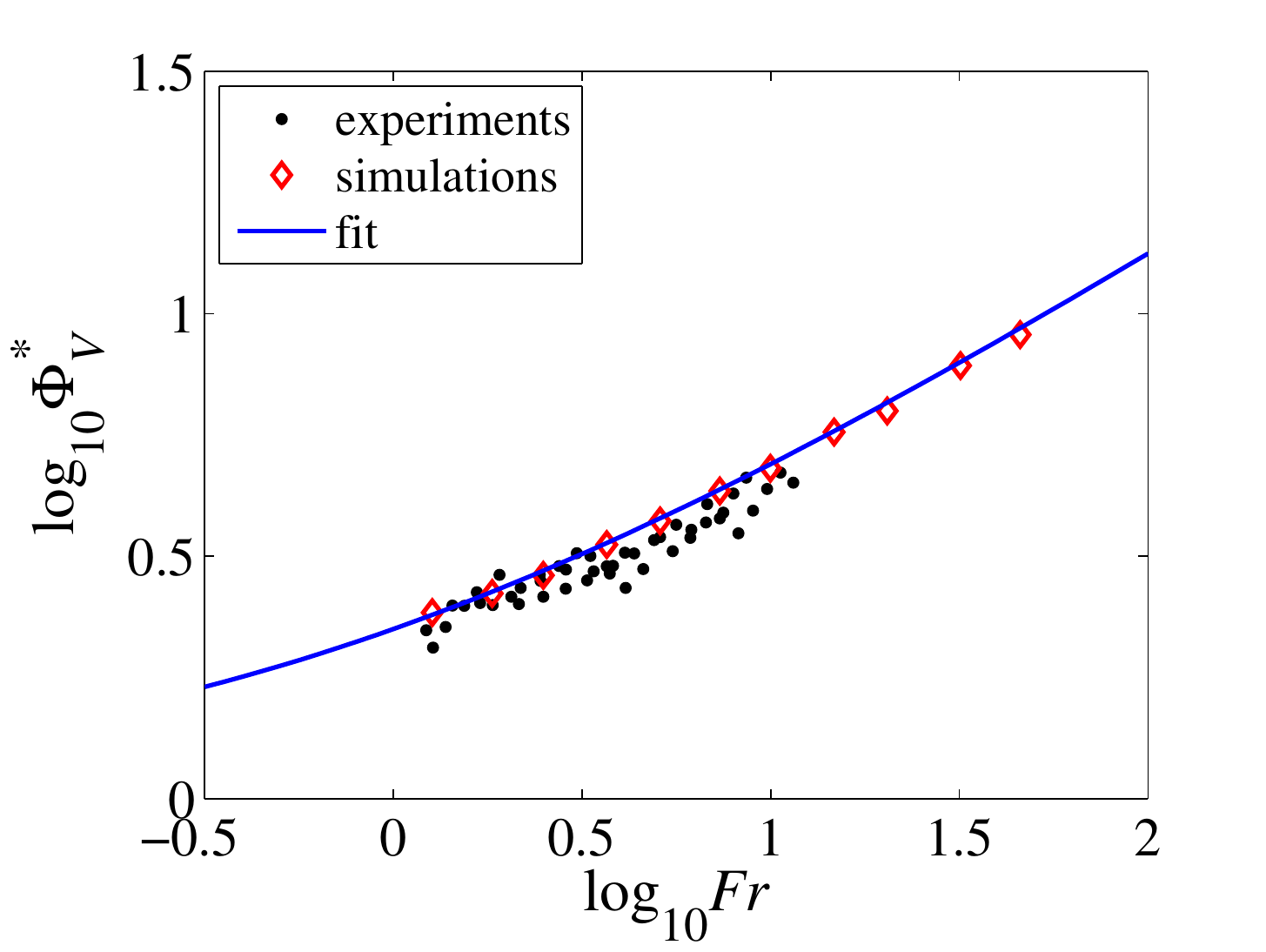}
    \caption{\label{fig:04} Flow rate calculated from the volume changes as a function of the Froude number in a double logarithmic plot. Both the experimental data (black dots) and the numerical data (red diamonds) correspond to the maximum value of $\Phi_V^*$. The range of experimental data is limited to $\Froude\approx12$ by the appearance of a surface seal. The blue line represents the fit
    $\Phi_V^*= 1.23 \Froude^{1/2} + 1.01$.}
\end{figure}
From Fig.~\ref{fig:03} we infer that in approach of the pinch-off point this flow rate becomes constant and can be determined as the maximum slope of the volume as a function of time (green line, Fig.~\ref{fig:03}), i.e., the air flow rate through the neck equals the rate of change of the volume of the cavity below the pinch-off depth, of course under the assumption that the gas flow remains in the incompressible limit. We subsequently non-dimensionalize this air flow rate $\Phi_V \equiv - dV/dt$ with the disk radius and the impact speed, namely, $\Phi_V^* \equiv \Phi_V/(R_0^2U_0)$, where the asterisk denotes a dimensionless value. We determined the flow rate for a number of different disk radii (ranging from $15$ to $30$ mm) and impact speeds ($0.45$-$1.30$ m/s), the results of which are shown in Fig.~\ref{fig:04} where we plot the dimensionless flow rate $\Phi_V^*$ versus the Froude number \Froude\ on a double-logarithmic scale (black dots). The experimental range is limited by the appearance of a surface seal at high impact speeds, where the crown splash is pulled inwards due to the air flow induced by the disc and closes the cavity at the surface. This surface seal usually has a significant influence on the cavity shape and dynamics~\cite{Bergmann2009} as well as the gas flow rate in the neck, so all of the experiments reported here are without surface seal. {Although the experimental data follows an apparent power-law, extending the experimental range in \Froude\ with our boundary integral code~\cite{Bergmann2009, Gekle2010a} reveals that the results do not lie on a straight line (Fig.~\ref{fig:04}, open red diamonds). This suggests that there does not exist a pure power-law.}

\bigskip
\emph{An analytical argument} -
Using the assumption that the cavity expansion and collapse take place in horizontal non-interacting layers of fluid --an assumption that was successfully used in Bergmann \emph{et al.}~\cite{Bergmann2009}-- we will now shed light onto the behavior of the air flow rate through the neck as a function of the Froude number. {We will provide an analytical argument in this subsection, which we work out in detail based on the model of Bergmann \emph{et al.}~\cite{Bergmann2009} in Appendix~\ref{app:ScalingLaw}.} For convenience from hereon we will take the $z$ to mean the depth below the undisturbed water surface, i.e., $z=0$ at the latter and increases with depth.

The quantity that we aim to calculate is the time rate of change of the cavity volume $\dot{V} = dV/dt$, i.e.,
\begin{equation}
    \label{eq:defPhi}
    \Phi_V \equiv -\frac{dV}{dt} = - \frac{d}{dt}\int_{z_c}^{z_\mathrm{disc}} 
    \!\!\!\!\!\pi\, [r(z,t)]^2 dz \,,
\end{equation}
where it is understood that the expression needs to be evaluated at the pinch-off time. Here, $r(z,t)$ is the cavity profile, $z_\mathrm{disc}(t)$ is the vertical position of the disc and $z_c$ is the pinch-off depth. We obtain
\begin{equation}
    \label{eq:volder}
    \Phi_V = - \int_{z_c}^{z_\mathrm{disc}} \!\!\!\!\!2\pi\, r(z,t)\,\dot{r}(z,t,) dz 
    \,\,\,-\,\,\, \pi R_0^2 U_0\, ,
\end{equation}
where the last term is due to the downward moving disc and $\dot{r} \equiv \partial r/\partial t$ denotes the radial velocity of the cavity wall.

To approximate the integral in Eq.~(\ref{eq:volder}) we subdivide the expanding and collapsing cavity --at times close to the collapse--  into regions {that we can express analytically using the model provided by Bergmann \textit{et al.}\cite{Bergmann2009}. This is done in Appendix~\ref{app:ScalingLaw} and leads to}

\begin{equation}
    \label{eq:flowscaling}
    \Phi_V = R_0^2U_0 \left[A \,\Froude^{1/2} + B \right] \,\,\mathrm{or}\, \Phi_V^* = A\, \Froude^{1/2} + B \,,
\end{equation}
with $A$ and $B$ numerical constants. To test this relation we extended the experiments of Fig.~\ref{fig:04} by performing boundary integral numerical simulations in order to cover a wide range of Froude numbers \footnote{The simulations in Fig.~\ref {fig:04} are two-phase boundary integral simulations, where close to pinch off the compressibility of the gas is taken into account using the one-dimensional compressible Euler equations (briefly discussed in Section~\ref {sec:Compressibility} as type (iii) simulations). More details about these simulations can be found in~\cite {Gekle2010,Gekle2010a}.}. 
The obtained results are added to Fig.~\ref{fig:04} using red diamonds. There is a good agreement with the experimental data, and the non-constant slope is clearly visible. A fit to the simulation data confirms Eq.~(\ref{eq:flowscaling}) and gives $A \approx 1.23$ and $B \approx 1.01$.

\section{Flow visualization}\label{sec:Visualization}
\begin{figure}
    \centering
    \includegraphics[width=8cm]{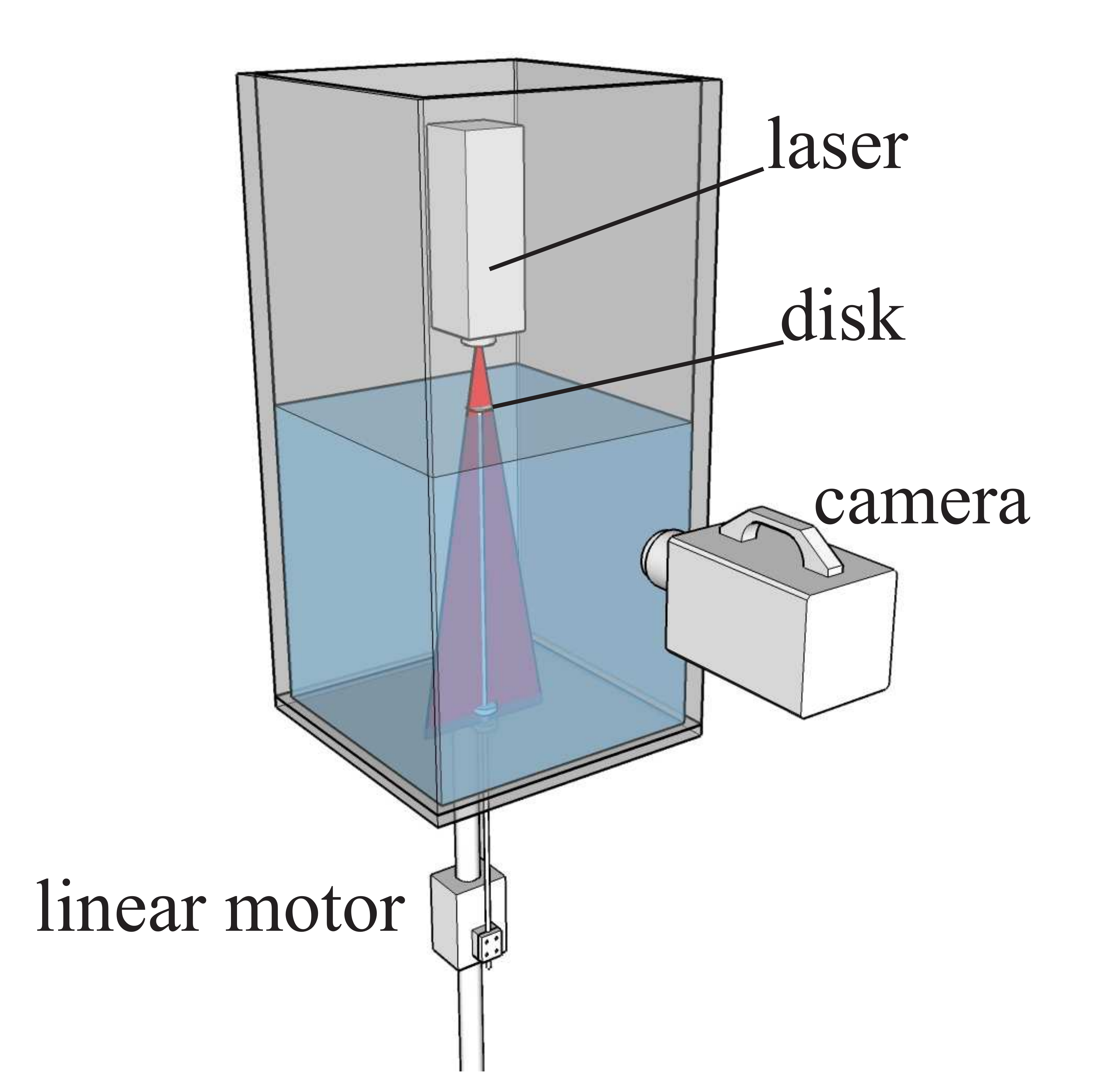}
    \caption{\label{fig:05} A schematic view of the setup. A laser sheet shines from above on the disk, illuminating the interior of the cavity after the disk has impacted the water surface. We insert smoke in the top part of the container and when the linear motor pulls the disk through the water surface at a constant speed, the smoke is entrained into the cavity.}
\end{figure}
In this Section we perform a direct determination of the air flow velocity by seeding the air with smoke and illuminating with a laser sheet {(see also Gekle \textit{et al.}\cite{Gekle2010})}, the results of which we will subsequently compare to the velocities that were determined indirectly and independently using volume measurements. We will first describe the method and results of the flow visualization that we used to measure the air flow inside the cavity. Before doing the impact experiment we fill the atmosphere above the water surface with small smoke particles. When subsequently the disk is moved down through the water surface, the smoke is dragged along, and fills the cavity created below the surface. We illuminate a thin sheet of the smoke using a $1500~\text{mW}$ diode laser line generator (Magnum II) and record the experiment at a recording rate up to $15~\text{kHz}$ by placing the high speed camera perpendicular to the laser sheet (Fig.~\ref{fig:05}). The smoke consists of small glycerine-based droplets (diameter $\sim3~\mu\text{m}$), produced by a commercially available smoke machine built for light effects in discotheques. A simple analysis shows that the particles are light enough to neglect all inertial effects at least in the range of accelerations that we can measure experimentally: At a velocity difference of $10~\mathrm{m/s}$ the Reynolds number is $\sim2$, meaning that we can assume Stokes drag. Knowing the force on the particle as a function of the velocity difference and the mass of the particles, we can calculate the movement of the droplets in an accelerating flow. We find that the particles follow the flow up to $25~\mathrm{m/s}$ with a velocity lag less than 2\%.

\bigskip
\emph{Correlation technique} -
\begin{figure}
    \centering
    \includegraphics[width=5cm]{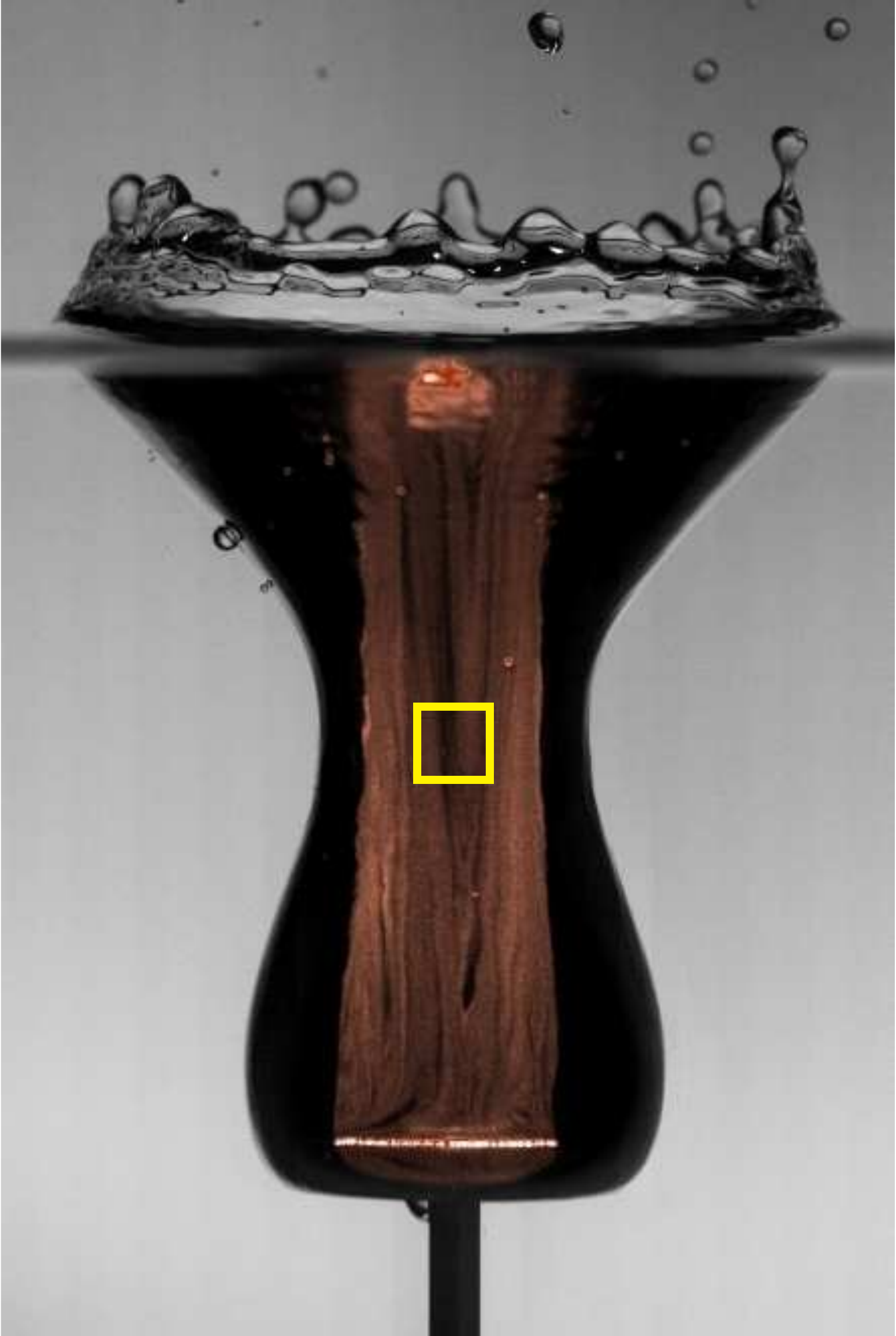}
    \caption{\label{fig:06} A snapshot of the cavity with an overlay of a recording of the illuminated smoke. The smoke particles are artificially colored orange in this figure. The size and position of the employed correlation window is indicated by the yellow square.}
\end{figure}
We determine the speed of the air in the neck by applying an image correlation velocimetry (ICV) technique~\cite{Tokumaru1995}. ICV differs from Particle Image Velocimetry (PIV) in the sense that we do not resolve discrete particles in our images, but we correlate smoke patterns instead of smoke particles. Figure~\ref{fig:06} shows the cavity with the illuminated smoke as an overlay, where the smoke is colored orange artificially for clarity. The actual measurements are done on a closer view of the cavity. The correlation is performed on a square correlation window, indicated by the yellow square. The width of the correlation window is $160$ pixels, corresponding to $8.8~\text{mm}$. In the latest stages we switch to a correlation window of $96$ pixels ($5.3~\text{mm}$) wide, anticipating for the smaller neck radius. The measurements are insensitive to small changes in the shape, size or position of the correlation window. The size of the window is optimized for quality of the cross correlation.

\begin{figure}
    \centering
    \includegraphics[width=8cm]{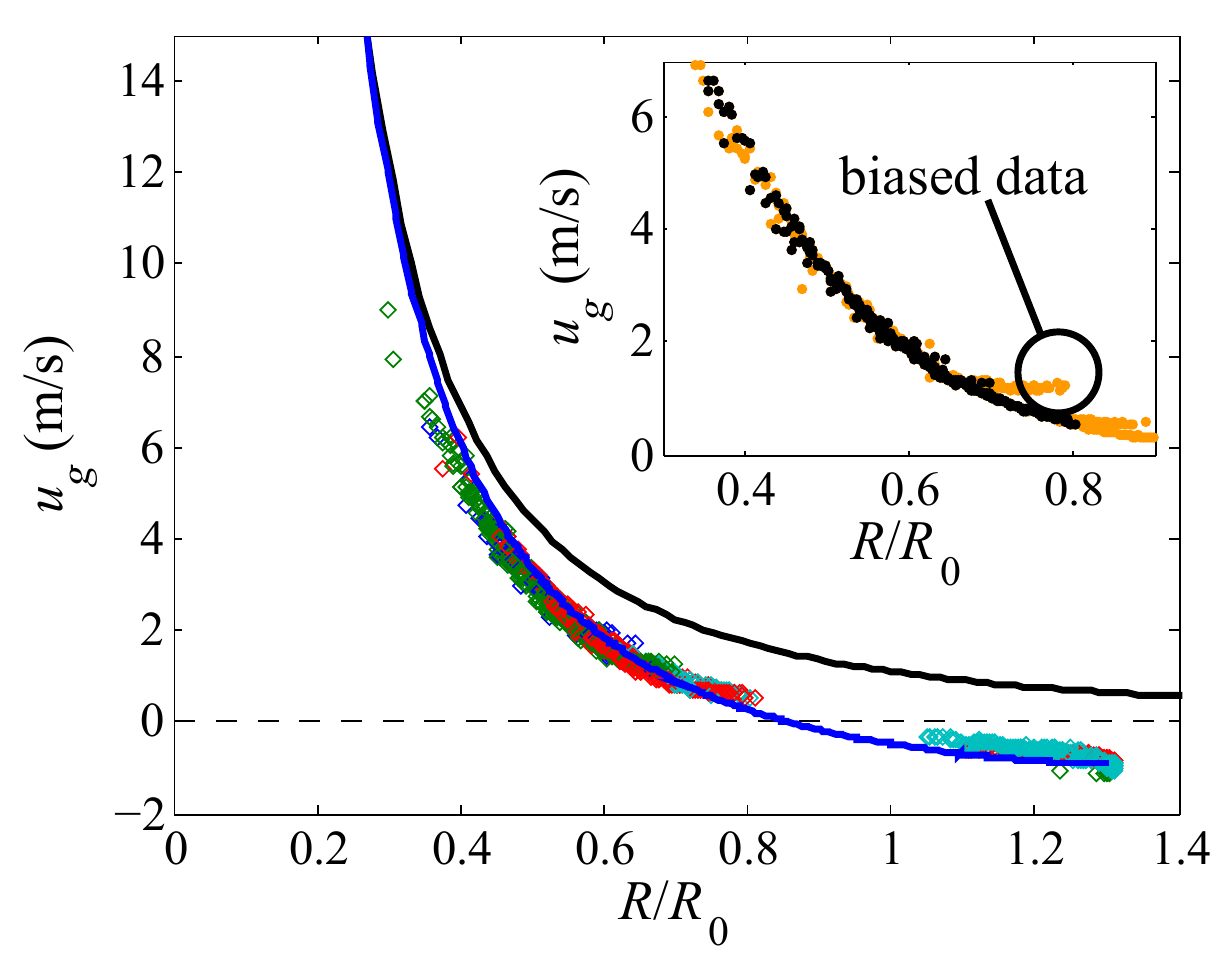}
    \caption{\label{fig:09}
    The vertical air velocity through the neck as a function of the neck radius $R$, measured in an experiment with $\Froude=5.1$ in three different ways: 
    (i) Directly, using smoke particles (diamonds), (ii) indirectly, using a smoothing polynomial fit to bubble volume of Fig.~\ref{fig:03} (blue line) , and (iii) indirectly, using a constant flow rate approximation, determined at pinch-off (cf. Fig~\ref{fig:03}, black line). The different colors of the diamonds correspond to different numbers of frames that are skipped in the cross-correlation (see main text). The inset shows the same vertical velocity data measured using method (i) for two different values for the peak-to-peak ratio $\lambda$: For $\lambda>1.5$ (orange dots) we find strongly biased data, which are eliminated using a higher threshold ($\lambda>3.5$, black dots).}
\end{figure}

To improve the ICV analysis we used an image subtraction technique that we describe in Appendix~\ref{app:ImageSubtraction}. A subtraction technique similar to the one that we use here has been used previously for double-frame PIV images~\cite{Honkanen2005}, where it was found that if the displacement of the particles is too small between a pair of images, the displacement peak in the correlation is biased. This bias is related to the particle size in pixels and the displacement in pixels. In our case this length scale does not exist because we do not resolve separate smoke particles in our experimental setup. Instead of calculating the expected bias, we identify biased values by their departure from the global trend of the data (Fig.~\ref{fig:09}, inset). As a remedy for the bias, we artificially increase the displacement by skipping frames. The smaller the velocity, the larger the number of frames we skip. In addition to this we note that the bias is less pronounced compared to the case in~\cite{Honkanen2005} because we construct the image pair from three images instead of two.

The biased data and other spurious data is removed by making an objective selection based on the peak-to-peak ratio of the correlation. This ratio is defined as the ratio between the two highest peaks in the correlation: $\lambda=p_1/p_2$. The inset of Figure~\ref{fig:09} shows the effectiveness of this selection method. We set $\lambda$ to values between 3.5 and 5.0, depending on the specific measurement, so that almost all spurious data is removed. Taking higher values for $\lambda$ removes too many valid data points; lower values allow for too many biased data points.

In Fig.~\ref{fig:09} we compare the air speed that we measured directly using smoke particles with the air velocity that we calculated indirectly using the change in volume of the cavity, as discussed in the previous Section. The air speeds are plotted versus the neck radius $R(t)$ at pinch-off depth instead of time; time increases from right to left in the figure, i.e., towards smaller values of $R$. The blue line is obtained using a polynomial (smoothing) fit to the volume-time data of Fig.~\ref{fig:03}, determining the flow rate $\Phi_V(t)$ from the time derivative of this fit [Eq.~(\ref{eq:defPhi})], and finally dividing by $\pi R(t)^2$ to obtain the velocity. We find a very good agreement between the direct (smoke) and the indirect (volume) measurements.

Finally, the black line in Fig.~\ref{fig:09} is obtained by setting the flow rate to a constant value, namely to that corresponding to the time derivative of the volume curve just before pinch off (the green line in Fig.~\ref{fig:03}). We observe that at early times (large $R$) there are large deviations from the other two datasets. This stands to reason, since at these times we are still far away from the pinch-off moment, and the gas flow rate in the neck has not yet become (approximately) constant. Close to pinch off however, for $R/R_0\lesssim0.4$, we find that the constant flow rate approximation and the smoothing fit both provide the same air speed.

\section{The role of compressibility}
\label{sec:Compressibility}

The fact that the air flow rate becomes constant together with the surface area of the neck becoming vanishingly small suggests that the velocity in the neck diverges towards pinch off. However, as was mentioned in the introduction, a real singularity of the air flow velocity is prevented by compressibility effects. In a previous publication we presented a directly visible effect of the compressed gas flow, namely the upwards motion of the position of the minimum neck radius~\cite{Gekle2010}. This upwards motion was seen both in experiments and in simulations that take the compressibility into account, and is absent in simulations that neglect compressibility. In the same paper we reported that, next to this upward motion of the neck, the extremely fast airflow affects the smoothness of the neck. Especially this last effect is important, since it is in contradiction with the assumptions in theoretical pinch-off models where the neck is assumed to be slender~\cite{Eggers2007,Gekle2009}.

The question that we intend to answer in the present Section is how the effects of compressibility show up in the measurement of the cavity volume and the air flow rate that can be deduced from it, as was presented in Section~\ref{sec:Geometric} of this work. More specifically we will investigate the position of the stagnation point of the flow in the cavity (see below) and the air flow rate towards the pinch-off moment. Following the method we used in~\cite{Gekle2010}, we will compare our experimental results with three different types of boundary integral simulations: (i) a single phase version, in which only the water phase is resolved, (ii) a two-phase version where both the liquid and the gas flow are resolved as incompressible inviscid media, and (iii) a compressible gas version where the compressibility of the gas phase is taken into account by substituting the incompressible axisymmetric gas phase equations by one-dimensional compressible Euler equations at that moment during the collapse when compressibility effects start to become significant. More details about the numerical method can be found in~\cite{Gekle2010,Gekle2010a}.

\begin{figure}
    \centering
    \includegraphics[width=8cm]{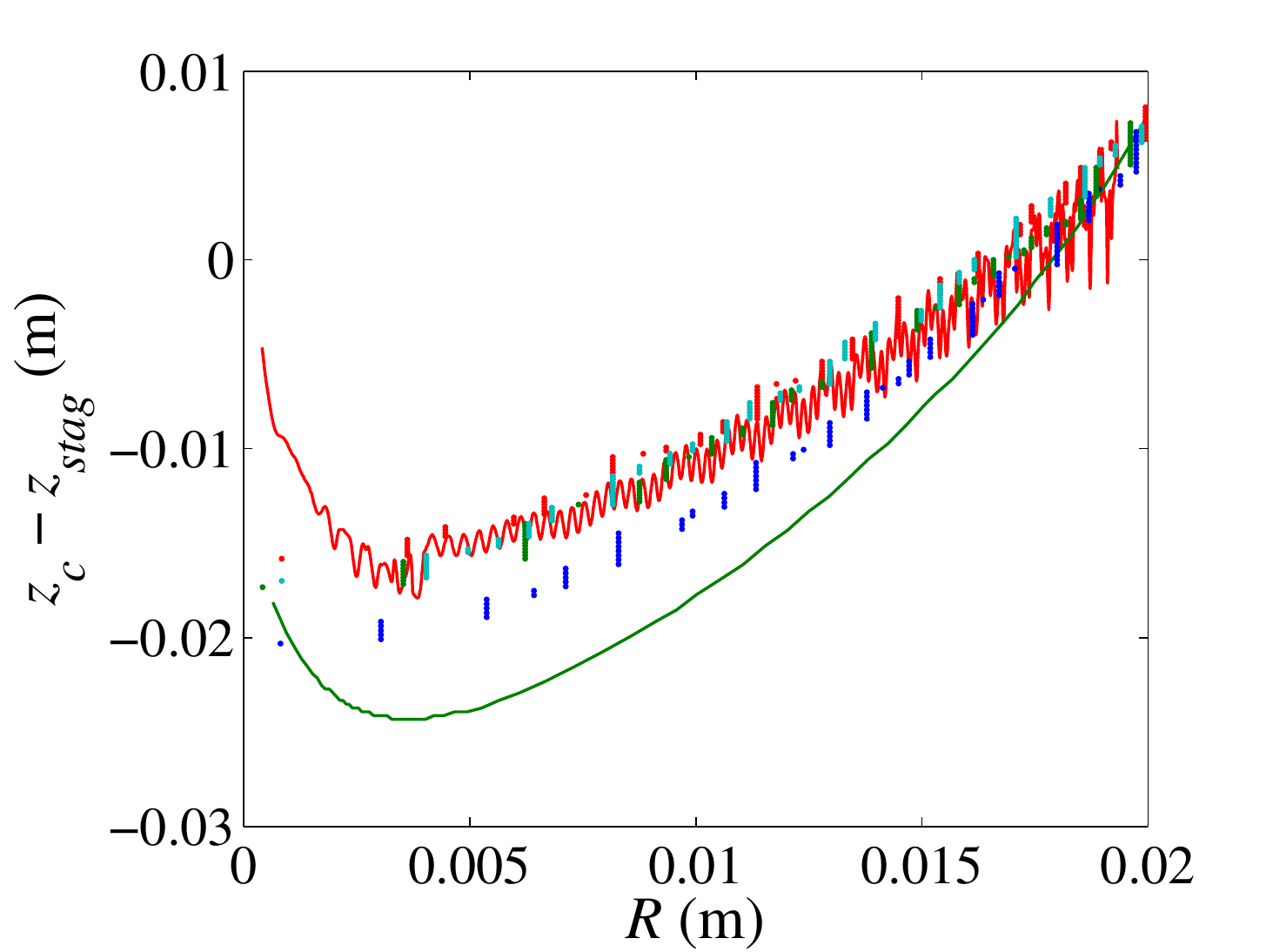}
    \caption{\label{fig:10}
    The location of the stagnation point $z_\mathrm{stag}$ with respect to that of the pinch-off point $z_{c}$ as a function of the neck radius $R$.
    Note that when that the stagnation point lies below the pinch-off point, $z_c-z_\mathrm{stag}$ is negative. Time increases from right to left (decreasing $R$). The dots are experimental data, obtained by volume measurements of four different experiments, where each color corresponds to a different experiment. All experiments were performed with disk radius $R_0=2.0~\text{cm}$ and impact speed $U_0=1.0~\text{m/s}$, i.e., $\Froude = 5.1$. The green line is the result of a two-phase boundary integral simulation without taking compressibility into account [type (ii)]. The red line is obtained by a two-phase boundary integral simulation which includes a compressible gas phase [type (iii)]. {Oscillations in the red line are a numerical artifact due to wave reflections in the compressible domain, see Gekle \textit{et al.}\cite{Gekle2010a} for details.}}
\end{figure}

\bigskip
\emph{Stagnation point} - Just above the disc the air must move downwards at approximately the same speed as the disc, whereas  simultaneously, towards closure, the air in the neck is moving upwards. This implies that somewhere in between there will be a stagnation point. We will estimate the location of this stagnation point as follows: The first step is to extend the analysis of Section~\ref{sec:Geometric}, where we tracked the volume below the pinch-off depth in time, to any depth $z$ below the pinch-off point. For every depth $z$ this will provide us with a curve similar to that in Fig.~\ref{fig:03} and by determining the time coordinate of the maximum we find the time $t_\mathrm{stag}$ at which the averaged \footnote{Averaged over the cross-sectional area of the cavity at that depth.} flow rate ($\sim \dot{V}$) at that depth $z \equiv z_\mathrm{stag}$ is zero. This point we then interpret as the location of the stagnation point $z_\mathrm{stag}(t_\mathrm{stag})$, which involves the assumption that close to the pinch-off moment the flow in the neck region becomes predominantly homogeneous and vertical. In Fig.~\ref{fig:10} we plot the measured location of $z_\mathrm{stag}$ for three different realizations of an experiment with a radius of 2 cm and an impact speed of 1 m/s. When we compare the experiments to a two-phase incompressible boundary integral simulation [type (ii)] (green line in Fig.~\ref{fig:10}), we find a considerable discrepancy between the two for small values of the neck radius $R$. If we however use the compressible version of the simulation [type (iii)], the agreement becomes much better (red line in Fig.~\ref{fig:10}{, see also Gekle \textit{et al.}\cite{Gekle2010}}), confirming the importance of compressibility in this limit. Note that experiments and both simulations do converge for larger values of $R$, where compressibility effects play no role.

The agreement is not perfect, however, which can partly be traced back to the technical difficulty of obtaining reliable values for $z_\mathrm{stag}$ from the experiment (which reflects in the large spread between the three different realizations) and partly to the fact that its determination neglects compressibility in a subtle way: Although in the experimental data compressibility is of course necessarily reflected in the shape of the cavity, the method of obtaining the air flow rate from it (namely by determining the time rate of change of the cavity volume) neglects compressibility in the air phase. A difference with the actual location of the stagnation point is therefore expected for high gas velocities (i.e., small neck radii).

\begin{figure}
    \centering
    \includegraphics[width=\textwidth]{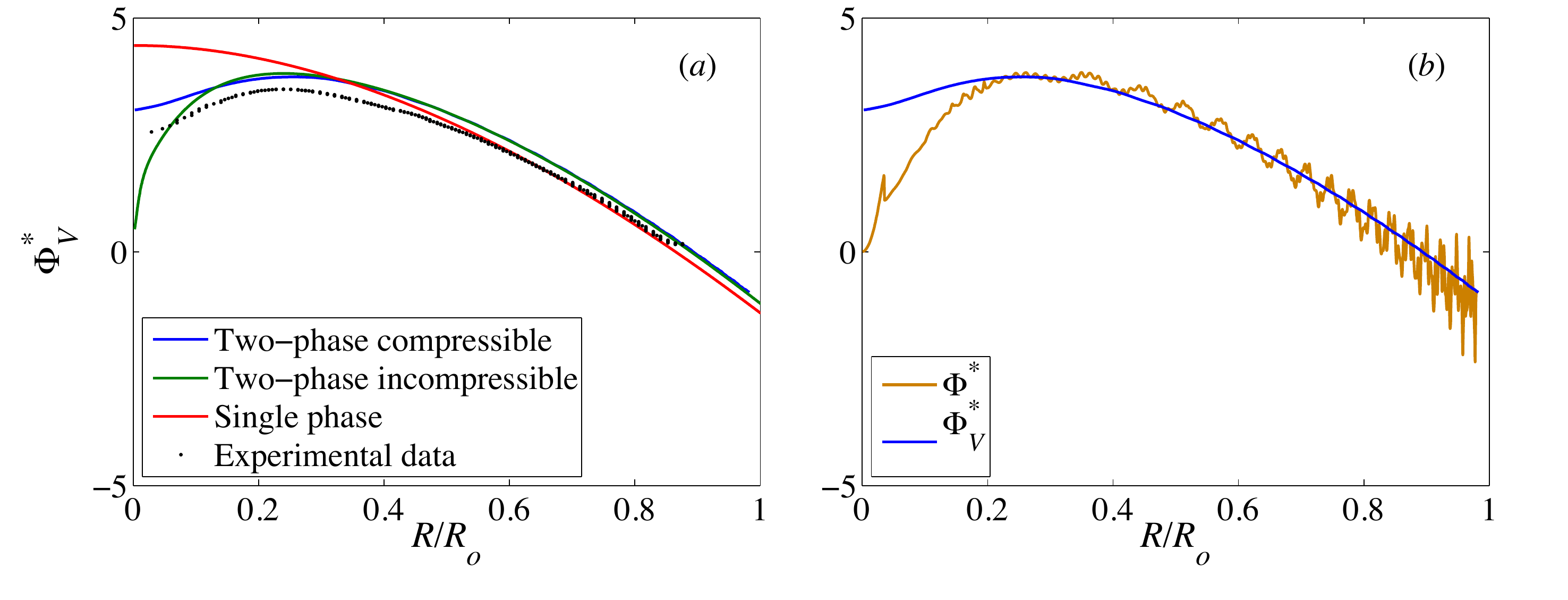}
    \caption{\label{fig:12AB}
    (a) The dimensionless \psd air flow rate $\Phi_V^* = \Phi_V/(R_0^2U_0)$ (from the time rate of change of the cavity volume)  as a function of the dimensionless neck radius $R/R_0$ in an impact experiment with disc radius $R_0 = 2~\text{cm}$ and impact speed $U_0=1~\text{m/s}$ ($\Froude = 5.1$). The black dots represent experimental data. The red line is obtained using a one-phase simulation [type(i)], which excludes the air phase. The green line is a two-phase boundary integral simulation without compressibility [type (ii)]. Finally, the blue line is the result of a two-phase boundary integral simulation which includes a compressible gas phase [type (iii)]. (b) Comparison of the dimensionless \psd air flow rate $\Phi_V^*$ [blue line; the same curve as in (a)] and the \tru air flow rate $\Phi^*$, both plotted versus $R/R_0$. The two curves diverge from each other below $R/R_0 \approx 0.2$.
    }
\end{figure}

\bigskip
\emph{Air flow rate} - As explained in the introduction of this Section, we can compare the air flow rate in the neck in experiment and simulation directly, by comparing the experimental velocities (cf. Section~\ref{sec:Visualization}) as was done in~\cite{Gekle2010}, but also indirectly, by using the volume analysis of Section~\ref{sec:Geometric} both in experiment and simulation. This second method, the results of which will be presented now, enables us to distinguish the effect the compressibility of the air has on the cavity wall (which is included in the analysis) from the pure compressibility of the flow (which is not included).

To do so, it is convenient to from now on distinguish the \tru air flow rate from the \psd air flow rate, i.e., the one obtained from the time rate of change of the cavity volume. In Fig.~\ref{fig:12AB}($a$) we plot the non-dimensionalized experimental \psd air flow rate in the neck,
\begin{equation}
    \label{eq:phiV}
	\Phi_V^* \equiv \frac{\Phi_V}{R_0^2U_0} \equiv -\frac{1}{R_0^2U_0}\frac{dV}{dt}\,,
\end{equation}
as a function of the dimensionless neck radius $R(t)/R_0$ (black dots), again for $\Froude = 5.1$. Repeating the experiment results in an uncertainty in the magnitude of $\Phi_V^*$ (corresponding to the spread of the experimental data in Fig.~\ref{fig:04}), but the behavior as a function of time is always the same: The \psd air flow rate in the neck reaches a maximum, and approaches a finite value towards the moment of pinch-off. We compare this result with those of the three different types of boundary integral simulations:

The one-phase code [type (i)] predicts a steadily increasing \psd air flow rate, which seems to level off to a constant value towards pinch off ($R/R_0 \to 0$). This is the red line in Fig.~\ref{fig:12AB}($a$).

The two-phase incompressible version [type (ii)] predicts a maximum at a location which is reasonably comparable to the experimental one, but after that decreases toward zero at the pinch-off moment (the green curve in Fig.~\ref{fig:12AB}($a$)). Since both phases are incompressible, this stands to reason: The pressure in the cavity rises instantly because of the divergence of the air velocity $u_g$ in the shrinking neck. This pressure decelerates the cavity wall, which in turn decreases the \psd air flow rate, which should go to zero in the $R/R_0 \to 0$-limit: In the context of incompressible flow, a finite \psd air flow rate would result in an infinite air velocity in the neck and consequently an infinite pressure within the cavity. Here it is good to note that for this two-phase incompressible code the \psd and \tru air flow rates are actually identical, due to the incompressibility of the air phase.

The two-phase compressible simulation [type (iii)]  also predicts a maximum for $\Phi_V^*$, at a location similar to the two-phase incompressible code and the experiment, but then decreases to a finite value for $R/R_0 \to 0$, just like the experiment. Clearly, and in contrast with the other two versions of the simulation which behave poorly, the agreement with the experiments is qualitatively very good and quantitatively satisfactory. All three types of simulations and the experiments all converge for larger $R/R_0 \approx \mathcal{O}(1)$, which is expected since airflow effects (let alone compressibility of the air phase) are small or even negligible in that regime.

\bigskip
The final question that we want to address is the difference between the \tru air flow rate (which incorporates all compressibility effects) and the \psd one (which only includes the effects of compressibility on the cavity wall). In experiment it is impossible to obtain the first quantity at the required precision, because its determination includes measurement errors in both air velocity $u_g$ and neck radius $R$. The two-phase compressible simulation technique however does offer a way to look at this difference: In Fig.~\ref{fig:12AB}($b$) we compare the \psd air flow rate $\Phi_V^*$ (the same curve as the blue one in Fig.~\ref{fig:12AB}($a$)) to the \tru air flow rate $\Phi^*$ (green curve), which is calculated from $u_g(t)$ and $R(t)$ as
\begin{equation}
    \label{eq:phi}
	\Phi^* \equiv \frac{\Phi}{R_0^2U_0}=
    \frac{1}{R_0^2U_0}{\pi u_g(t)R(t)^2}\,,
\end{equation}
both as a function of the dimensionless cavity radius $R(t)/R_0$. Clearly the two curves coincide above $R/R_0\approx 0.2$, but start to depart from one another below this value, indicating that here the compressibility of the air itself becomes significant, in good agreement with what we concluded from the previous plot (Fig.~\ref{fig:12AB}($a$)). We observe that the \tru air flow rate goes to zero for $R/R_0 \to 0$ (and incidentally not quite unlike the two-phase incompressible curve (green) in Fig.~\ref{fig:12AB}($a$)). This is of course what should happen, since the gas velocity in the neck needs to remain finite at all times. The difference between the two curves is {related to} the rate at which the gas in the cavity is compressed.

\section{Conclusions}
We have measured the air flow inside the neck of a collapsing cavity that was created by the impact of a circular disc on a water surface. More specifically we have performed and compared two types of experiments: First, we did indirect measurements, using the time rate of change of the cavity volume as a measure for the air flow rate in the neck, thereby neglecting compressibility of the air inside the cavity. Secondly, we performed direct measurements of the velocity in the neck of the cavity using image correlation velocimetry. Numerical boundary integral simulations of three different types have been used to evaluate and discuss our experimental findings.

For the complete experimentally available range of Froude numbers we showed that there is a very good agreement between the indirectly measured air flow rate and the boundary integral simulations. With the simulations we were able to extend the range of experimentally attainable Froude numbers, which revealed that the air flow rate is not a pure power-law of the Froude number. We formulated an analytical argument revealing that the dimensionless air flow rate should scale as $A \Froude^{1/2} + B$. Such a scaling compares well with experiments and simulations for $A \approx 1.23$ and $B \approx 1.01$.

By performing careful image correlation velocimetry experiments with a smoke-filled cavity we have been able to directly measure the air flow for relatively low air speeds, corresponding to $R/R_0 \geq 0.3$. In this region we found excellent agreement with the gas velocities that we calculated from the indirect measurements of the air flow rate and the neck radius $R(t)$.

Due to the very high air speed close to the moment of pinch-off ($R/R_0 \leq 0.2$) compressibility of the air can not be neglected anymore. We have demonstrated this by comparing experimental results to three types of numerical simulations: (i) one-phase boundary integral simulations, (ii) two-phase boundary integral simulations with an incompressible gas-phase, and (iii) a compressible gas version of the second type of simulations that include the gas phase as a compressible fluid. We analyze the time evolution of both the location of the stagnation point in the gas flow and the \psd air flow rate and explain our experimental observations in terms of the three types of simulations. The main conclusion is that the behavior that we observe in the experiments can only be reproduced by the simulations if compressibility is taken into account.

\appendix
\section{Derivation of the scaling law for $\Phi_V^*$}
\label{app:ScalingLaw}

\begin{figure}
    \centering
    \includegraphics[width=6cm]{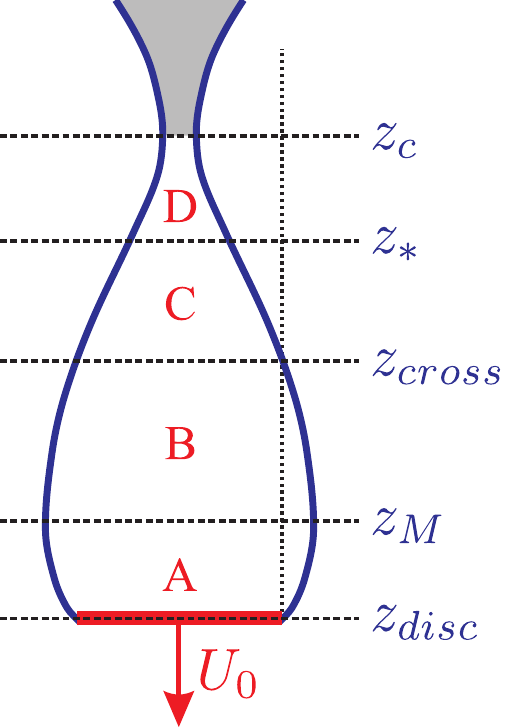}
    \caption{ \label{fig:app} {For the derivation in this Appendix, the cavity close to the pinch-off moment needs be divided into four regions:} The expansion region (A), between the location of the disc $z_{disc}$ and the location of the maximum $z_M$, where the cavity expands against hydrostatic pressure; the contraction region (B), between $z_M$ and the point $z_{cross}$ where the cavity reaches the disc radius again where the hydrostatic pressure approximation [Eq.~(\ref{eq:A2})] is matched to the inertial approximation [Eq.~(\ref{eq:A4})]; the collapse region (C) between $z_{cross}$ and $z_*$, characterized by continuity; and the self-similarity region (D), between $z_*$ and the pinch off location $z_c$, which is in addition characterized by a coupling between the vertical and horizontal coordinates.
    }
\end{figure}

In this Appendix we {derive Eq.~(\ref{eq:flowscaling}), the main result of \S\,\ref{subsec:airflowrate},} starting from the description of the cavity proposed in~\cite{Bergmann2009}. The starting point is the two-dimensional Rayleigh equation for the cavity wall $r(z,t)$, which originates from integrating the Euler equations in uncoupled horizontal layers of flow from some far away point $R_\infty$ to the cavity wall
\begin{equation}
\label{eq:A1}
\log \left(r/R_\infty\right)\frac{d}{dt}\left(r\dot{r}\right)\,+\,\tfrac{1}{2}\dot{r}^2 = g z
\end{equation}
in which $\dot{r} = \partial r/\partial t$ and $g$ is the acceleration of gravity.

This equation is solved in two different limits to describe the different regions in Fig.~\ref{fig:app}. The first one is to describe region A and B, taking for every depth $z$ the moment $t_{M}(z)$ of maximum expansion as a reference point. With $r(t_{M}) = R_{M}(z)$, $\dot{r}(t_{M}) = 0$, we can neglect the second term in Eq.~(\ref{eq:A1}) and replace the slowly varying logarithm in the first term by a constant \footnote{Although strictly speaking $R_{M}(z)$ is a function of $z$, it is slowly varying and can approximated by a constant when the logarithm of this quantity is taken.} 
$\beta \equiv \log(R_{M}/R_\infty)$ \footnote{The constant $\beta $ is different in the expansion $\beta_{expa}$ and the contraction phase $\beta_{ctra}$.} 
and solve
\begin{equation}
\label{eq:A2}
r(z,t)^2 = R_M(z)^2 - \frac{g z}{\beta}(t-t_M(z))^2 \,.
\end{equation}
In \cite{Bergmann2009} it was shown that
\begin{equation}
\label{eq:A3}
t_M(z) = \frac{z}{U_0} + \alpha_{expa}\beta_{expa} \frac{R_0U_0}{gz} \,,
\end{equation}
in which the first term represents the time span needed to arrive at depth $z$ and the second the amount of time to expand to the maximum radius. Here $\alpha_{expa}\beta_{expa}$ is a constant \footnote{The nomenclature of the constants is chosen such as to be consistent with \cite {Bergmann2009}.}. 

The second approximate solution corresponds to the small $R$ limit in the collapse regions C and D of Fig.~\ref{fig:app}, in which both the driving pressure $gz$ and the inertial term $\dot{r}^2/2$ can be considered small when $|\log(r/R_\infty)| \gg 1/2$. This then leads to $\tfrac{d}{dt}(r\dot{r}) = 0$ which is readily solved to give:
\begin{equation}
\label{eq:A4}
r(z,t)^2 = 2\alpha_{ctra} R_0U_0 \left(t_{coll}(z) - t\right) \,,
\end{equation}
in which $\alpha_{ctra}$ is a constant and $t_{coll}(z)$ is the (virtual) closure time of the cavity at depth $z$.  At any depth $z$ the approximate solutions are tied together at the maximum (where a solution Eq.~(\ref{eq:A2}) with $\beta = \beta_{expa}$ is matched to a solution with $\beta=\beta_{ctra}$) and at the moment $t_{cross}(z)$ when $r(z,t) = R_0$ again. Here, the solution Eq.~(\ref{eq:A2}) with $\beta = \beta_{ctra}$ is matched to Eq.~(\ref{eq:A4}). More details can be found in \cite{Bergmann2009}.

The quantity we want to calculate is Eq.~(\ref{eq:volder}), which contains the time derivatives of Eqs.~(\ref{eq:A2}) and (\ref{eq:A4}), which are:
\begin{eqnarray}
	\label{eq:A5}
	\frac{d}{dt}\left(r(z,t)^2\right) &=& - 2\frac{g z}{\beta}\left(t-t_M(z)\right)\,\,\,\, \mathrm{(region A,B)}\nonumber\\
	\frac{d}{dt}\left(r(z,t)^2\right) &=& - 2\alpha_{ctra} R_0U_0\,\,\,\, \mathrm{(region C,D)}\,.
\end{eqnarray}
which subsequently need to be evaluated at the moment of pinch-off $t = t_c$, for which it was derived in \cite{Bergmann2009} that it is independent of the impact speed: $t_c = C_2 \sqrt{R_0/g}$ \footnote{The constant $C_2$ is not independent of the $\alpha$'s and $\beta$'s: $C_2 = 2 (\alpha_{expa}\beta_{expa}+\alpha_{ctra}\beta_{ctra})^{1/2}$.}.
Inserting this expression together with Eq.~(\ref{eq:A3}) into the first Eq.~(\ref{eq:A5}) gives
\begin{eqnarray}
	\label{eq:A6}
	&&\frac{d}{dt}\left(r(z,t_c)^2\right) =\\
	&&- 2\frac{g z}{\beta}\left(C_2\sqrt{\frac{R_0}{g}} - \frac{z}{U_0} - \alpha_{expa}\beta_{expa} \frac{R_0U_0}{gz}\right)\,. \nonumber
\end{eqnarray}
Finally, we need to integrate the second Eq.~(\ref{eq:A5}) and Eq.~(\ref{eq:A6}) over $z$ between $z_{disc}$ and $z_c$. This is a straightforward calculation which gives the following lengthy result
\begin{eqnarray}
	\label{eq:A7}
	&&\int_{z_c}^{z_\mathrm{disc}} \frac{d}{dt}\left(r(z,t_c)^2\right) dz=\\
	&&- \frac{C_2}{\beta_{expa}}\sqrt{R_0g}\left(z_\mathrm{disc}^2 - z_M^2\right) 
	+ \frac{2}{3\beta_{expa}} {\frac{g}{U_0}} \left(z_\mathrm{disc}^3 - z_M^3\right) 
	+\nonumber\\
	&&2\alpha_{expa}R_0U_0(z_\mathrm{disc}- z_M) 
	- \frac{C_2}{\beta_{ctra}}\sqrt{R_0g}\left(z_M^2 - z_\mathrm{cross}^2\right) 
	+ \nonumber\\
	&& \frac{2}{3\beta_{ctra}} {\frac{g}{U_0}} \left(z_M^3 - z_\mathrm{cross}^3\right) 
	+ 2{\frac{\alpha_{expa}\beta_{expa}}{\beta_{ctra}}}R_0U_0(z_M-z_\mathrm{cross}) 
	- \nonumber\\
	&&2\alpha_{ctra} {R_0U_0} (z_{cross}-z_*) 
	- 2\alpha_{ctra} {R_0U_0} (z_*-z_c) \,. \nonumber
\end{eqnarray}
We now use that all length scales $z_{disc}$, $z_M$, $z_{cross}$, and $z_*$ scale as $R_0 \Froude^{1/2}$, except for the difference $(z_*-z_c)$, which due to the self-similarity in the neck radius scales as $R_0$. This means that the above Eq.~(\ref{eq:A7}) has the folowing form
\begin{eqnarray}
	\label{eq:A8}
	&&\int_{z_c}^{z_\mathrm{disc}} \frac{d}{dt}\left(r(z,t_c)^2\right) dz=\\
	&& - \kappa_1 \sqrt{R_0g}\, R_0^2 \Froude + \kappa_2 \frac{g}{{U_0}} \,R_0^3 \Froude^{3/2} + \kappa_3 R_0^2U_0 \Froude^{1/2} - \kappa_4 R_0^2U_0\,,  \nonumber
\end{eqnarray}
in which $\kappa_1$-$\kappa_4$ are positive numerical constants, which depend on the $\alpha$'s, $\beta$'s and the proportionality constants in the scaling laws for the length scales $z_{disc}$, $z_M$, $z_{cross}$, $z_*$, and $(z_*-z_c)$ . By writing {$g=U_0^2R_0^{-1}\Froude^{-1}$} in the first two terms we finally obtain:
\begin{eqnarray}
	\label{eq:A9}
	&&\int_{z_c}^{z_\mathrm{disc}} \frac{d}{dt}\left(r(z,t_c)^2\right) dz=\qquad\qquad\\
	&&(- \kappa_1 + \kappa_2 + \kappa_3) \, R_0^2U_0\, \Froude^{1/2} \,\,- \,\,\kappa_4\, R_0^2U_0\,.  \nonumber
\end{eqnarray}
If we now insert the above result in Eq.~(\ref{eq:volder}) we obtain
\begin{eqnarray}
\label{eq:A10}
\Phi_V &=& - \int_{z_c}^{z_\mathrm{disc}} \!\!\!\!\!2\pi\, r(z,t)\,\dot{r}(z,t,) dz \,\,\,-\,\,\, \pi R_0^2 U_0 \\
&=& \pi (\kappa_1 - \kappa_2 - \kappa_3) \, R_0^2U_0\, \Froude^{1/2} + \pi(\kappa_4-1)R_0^2 U_0 \nonumber
\end{eqnarray}
which then leads to
\begin{equation}
\label{eq:A11}
\Phi_V^* \equiv \frac{\Phi_V}{R_0^2U_0} = A\, \Froude^{1/2} + B \,,
\end{equation}
with $A \equiv  \pi (\kappa_1 - \kappa_2 - \kappa_3)$ and $B \equiv  \pi(\kappa_4-1)$. 

\section{Image subtraction technique}
\label{app:ImageSubtraction}
Mainly due to reflections from and refraction at the free surface, there are structures visible in the correlation window that move slowly compared to the typical gas velocities that we want to measure. A correlation between two unprocessed images gives a strong correlation peak close to zero because these structures are dominating the image, and thereupon also the cross-correlation. Standard background subtraction is not able to remove these features since, because of their refractive and reflective nature, they appear and disappear at unpredictable instances in time and are not stationary. Instead we use the difference between subsequent images in the following way: We start with three images $I_n$, $I_{n+1}$, and $I_{n+2}$. After applying a low pass filter we create two new images by mutual subtraction of the original three images: $J_n=I_{n+1}-I_n$ and $J_{n+1}=I_{n+2}-I_{n+1}$. We then apply a min-max filter~\cite{Westerweel1993} to these images $J_n$ and $J_{n+1}$, followed by the cross correlation of $J_n$ and $J_{n+1}$. On the result of the correlation we apply a multiple peak detection to find the highest peak $p_1$ and the second-highest peak $p_2$. We determine the position of the highest peak with sub-pixel accuracy by a gaussian fitting routine.

\begin{acknowledgments}
We acknowledge the Netherlands Organisation for Scientific Research (NWO) for financial support through the Spinoza program.
\end{acknowledgments}

\bibliography{airflow}

\end{document}